\documentclass[aps,showpacs,twocolumn]{revtex4}

\usepackage[dvips]{graphicx}
\usepackage{amsmath,array,amssymb}
\usepackage{amsfonts}
\usepackage{color}
\usepackage{natbib}
\usepackage{wrapfig}

\newcommand{\be}{\begin{equation*}}
\newcommand{\ee}{\end{equation*}}
\newcommand{\bea}{\begin{eqnarray*}}
\newcommand{\eea}{\end{eqnarray*}}

\begin{document}

\title{Comment on ``Phase separation in a two-species Bose mixture''}

\author{Fei Zhan and Ian P. McCulloch}

\affiliation{Centre for Engineered Quantum Systems, School of Mathematics and Physics,
The University of Queensland, St Lucia, QLD 4072, Australia}
\begin{abstract}
 In an article in 2007, Mishra, Pai, and Das [Phys. Rev. A {\bf 76}, 013604 (2007)] investigated the
two-component Bose-Hubbard model using the numerical DMRG procedure. In
the regime of inter-species repulsion $U^{ab}$ larger than the
intra-species repulsion $U$, they found a transition from a uniform
miscible phase to phase-separation occurring at a finite value of $U$, {\it e.g.}, at
around $U = 1.3$ for $\Delta = U^{ab}/U = 1.05$ and $\rho_{a}=\rho_{b}=1/2$. In this comment, we show
that this result is not correct and in fact the two-component Bose-Hubbard
model is unstable to phase-separation for any $U^{ab} > U >0$.
\end{abstract}

\maketitle

In an article in 2007, Mishra {\it et al.} \cite{mishra2007pra}
studied the two-component Bose-Hubbard model and phases that can be
described by this model with their modified form of the finite-size
density matrix renormalization group (FSDMRG) method.
The studied lattice has a density of $\rho_{a(b)}$ bosons of species $a(b)$ per site, with
intra-species repulsion $U$,
and inter-species repulsion $U^{ab}$
while the tunneling coefficient $t$ is chosen as the energy unit.
One of their results is that for any fixed $\Delta = U^{ab}/U > 1$ the system
undergoes a transition from a miscible phase at small $U$ to phase-separation at large $U$.
Unfortunately, we find this conclusion is invalid and the system is phase separated
whenever $\Delta>1$, for all values of $U>0$.

We show this by perturbation analysis, where the miscible phase is unstable at first
order for any density profile and $U>0$, $\Delta>1$.
As an example, we also perform iDMRG calculation for the density profile
$\rho_{a}=\rho_{b}=1/2$ with $\Delta=1.05$,
which is one of the three studied density profiles
in Ref.~\cite{mishra2007pra}.
Moreover, by finite DMRG simulation with the above set of parameters,
we point out that a plausible reason for the mistake of Mishra {\it et al} is
they have not done a sufficient number of sweeps in their finite-size DMRG algorithm.

When $U\ll t$,
we can prove by a first-order perturbation theory in the thermodynamic limit
that the phase-separated energy per site is always lower than that of the miscible phase.
The Hamiltonian is comprised of the kinetic term and the on-site repulsion term,
\begin{equation}
 H=H_{T}+H_{U}.
\end{equation}
On $L$ sites with periodic boundary conditions,
these terms take the form in the momentum space as
\begin{align}
 H_{T}=&-2t\sum_{q=0}^{L-1}\cos(2\pi q/L)(a_{q}^{\dagger}a_{q}+b_{q}^{\dagger}b_{q}),
\end{align}
\begin{align}
 H_{U}=&\frac{U}{2L}\sum_{\substack{q_{1},q_{2},\\q_{3},q_{4}\\=0}}^{L-1}
\delta_{q_{1}+q_{2},q_{3}+q_{4}}(a_{q_{1}}^{\dagger}a_{q_{2}}^{\dagger}a_{q_{3}}
a_{q_{4} }+b_{q_{1}}^{\dagger}b_{q_{2}}^{\dagger}b_{q_{3}}b_{q_{4}})\notag\\
 &+\frac{U^{ab}}{L}\sum_{\substack{q_{1},q_{2},\\q_{3},q_{4}\\=0}}^{L-1}
\delta_{q_{1}+q_{2},q_{3}+q_{4}}a_{q_{1}}^{\dagger}b_{q_{2}}^{\dagger}a_{q_{3}}b_{q_{4}}
,
\end{align}
where the operator in momentum space is defined as 
\begin{equation}
 a_{q}^{\dagger}(b_{q}^{\dagger})=\frac{1}{\sqrt{L}}\sum_{j=1}^{L}e^{i(j2\pi q/L)}a_{j}^{\dagger}(b_{j}^{\dagger})
\end{equation}
creating a species $a(b)$ boson with momentum $2\pi q/L$.

\begin{figure}[t]
 \includegraphics[width=.9\linewidth]{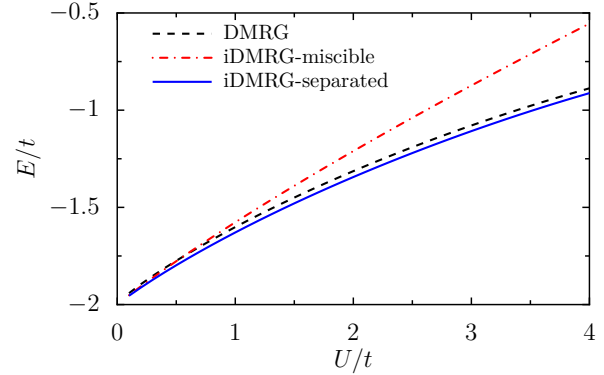}
\caption{\label{fig:idmrg} (Color online) The total energy per site obtained from finite DMRG (black dashed),
infinite DMRG for a miscible phase (red dash-dotted), and for phase-separation (blue solid).}
\end{figure}

In the ground state of the miscible phase with $U=0$
all the bosons are in the $q=0$ level,
therefore we have
\begin{equation}
 |{\rm miscible}\rangle=\frac{1}{\sqrt{n^{a}!n^{b}!}}(a_{q=0}^{\dagger})^{n^{a}}(b_{q=0}^{\dagger})^{n^{b}}|0\rangle,
\end{equation}
where $n^{a(b)} = \rho_{a(b)} L$ is the number of species $a(b)$ bosons.

In the phase-separated regime, the system will split into two domains,
each with momentum $q\rightarrow0$ in the thermodynamic limit.
Each domain only has one species present, therefore we can write
the wavefunction in this region, {\it e.g.,} with only species $a$ present, as
\begin{equation}
 |\text{phase-sep.}\rangle=\frac{1}{\sqrt{n^{a}!}}(a_{q=0}^{\dagger})^{n^{a}}|0\rangle,
\end{equation}
and similarly for species $b$.

Both states give the same kinetic energy per site, $E^{K}=-2t(\rho_{a}+\rho_{b})$.
But the first-order perturbation gives different corrections:
For the miscible state, we have
\begin{equation}
 E^{U}_{\text{miscible}} = \frac{U}{2}\left(\rho_{a}^{2}+\rho_{b}^{2}+2\Delta\rho_{a}\rho_{b}\right);
\end{equation}
On the other hand, we have
\begin{equation}
 E^{U}_{\text{phase-sep.}} = \frac{U}{2}\left(\rho_{a}+\rho_{b}\right)^{2}.
\end{equation}
In consequence, as long as $\Delta>1$,
the phase-separated energy per site is lower than the miscible phase.

\begin{figure}[t]
 \includegraphics[width=\linewidth]{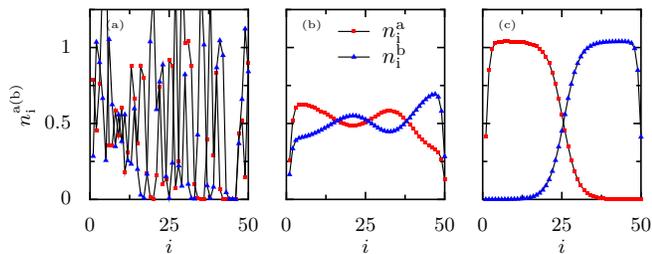}
\caption{\label{fig:evolve} (Color online) The expectations $n^{a}_{i}$ and $n^{b}_{i}$ for (a) an initial random wavefunction,
(b) the wavefunction after about $20$ sweeps, and (c) after about another $200$ sweeps when $U=1$.}
\end{figure}

The same conclusion can be drawn from the calculation of the ground state energy
per site of a one-component Bose-Hubbard model
using an infinite DMRG (iDMRG)~\cite{mcculloch08arxiv}.
A ground state with the density $\rho=1/2$ simulates a state in the miscible phase
and with the density $\rho=1$ it simulates a phase-separated state.

\begin{figure}
 \includegraphics[width=\linewidth]{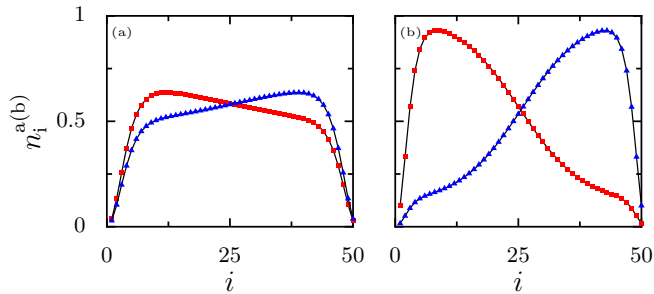}
\caption{\label{fig:smallu} (Color online) The expectations $n^{a}_{i}$ and $n^{b}_{i}$ when
(a) $U=0.1$ and (b) $U=0.2$, respectively (symbols used as in Fig.~\ref{fig:evolve}).}
\end{figure}

In Fig. \ref{fig:idmrg}, we compare the total energy per site for the miscible phase
and the  phase-separation.
We can find when $U$ is comparable to $t$ the miscible phase apparently has a higher energy than the phase-separation.
In addition, the finite DMRG gives a slightly higher total energy.
The tiny extra energy should stem mainly from the open boundary and from the domain wall between two domains in the phase-separated state.
We also verified that for $\Delta=1$, the DMRG calculation produces a miscible phase as expected.

When the energy
difference between miscisble and phase-separated states is small, it may
take a lot of iterations for DMRG to converge to the correct state.
For instance, we recognize the imbalance in occupations
near the boundaries in Fig.~3 of Ref.~\cite{mishra2007pra} with $U=1$ is a precursor
to a fully phase-separated state.
We have carried out DMRG
calculations for an example parameter set $\rho_{a}=\rho_{b}=1/2$, $\Delta=1.05$,
to verify that DMRG does reproduce the expected phase separated state.
This is shown in Fig.~\ref{fig:evolve}, where we start from a random wavefunction.
The randomness can be seen in the occupation expectations $n^{a}_{i}$ and $n^{b}_{i}$ in Fig. \ref{fig:evolve}(a).
After around $20$ sweeps, $n^{a}_{i}$ and $n^{b}_{i}$ evolve to a pattern displayed in Fig. \ref{fig:evolve}(b),
where we find the phase-separation also starts from the boundaries.
After another about $200$ sweeps, the occupation expectations are clearly phase-separated as shown in Fig. \ref{fig:evolve}(c).

In conclusion, we have shown through a perturbation analysis that the
two-species Bose mixture is unstable to phase separation whenever
$\Delta>1$, for any $U>0$. Additionally, We have also carried out DMRG
calculations for an example parameter set to verify that DMRG does
reproduce the expected phase-separated state.
We can easily see in Fig.~\ref{fig:smallu}(a) and (b) that
even when $U$ is very small, two species of bosons could not coexist and
two domains are formed when enough sweeps have been done.  Therefore, the erroneous
conclusion in Ref. \cite{mishra2007pra} is likely due to an insufficient number of sweeps.

We acknowledge the support from the Australian Research Council Centre of Excellence for
Engineered Quantum Systems and the Discovery Projects funding scheme (Project No. DP1092513).


\begin{thebibliography}{35}
\bibitem{mishra2007pra} T. Mishra, R. V. Pai, and B. P. Das, Phys. Rev. A {\bf 76}, 013604 (2007).
\bibitem{mcculloch08arxiv} I. P. McCulloch, arXiv:0804.2509 (2008).
\end{thebibliography}

\end{document}